\documentclass{article}
\usepackage{graphicx}

\input{tcilatex}
\begin{document}

\title{An inverter-chain link implementation of quantum teleportation and
superdense coding}
\author{Felix A. Buot,$^{1,2}$ Roland E.S. Otadoy,$^{1}$and Xavier L. Bacalla%
$^{1}$ \\
$^{1}$LCFMNN, TCSE Group, Department of Physics, \\
University of San Carlos, Cebu City, Philippines, and\\
$^{2}$C\&LB Research Institute, Carmen, Cebu 6005, Philippines}
\maketitle

\begin{abstract}
A new perspective in terms of inverter-chain link (ICL) diagrams of quantum
entanglement faithfully captures the fundamental concept of quantum
teleportation and superdense coding. The ICL may be considered a series of $%
\sigma _{x}$ Pauli-matrix operations, where a physical/geometric
representation provides the mysterious link raised by EPR. Here, we employ
discrete phase space and ICL analyses of quantum entanglement as a resource
for quantum teleportation and superdense coding. We underscore the quantum
superposition principle and Hadamard transformation under a local
single-qubit operation. On the fundamental question posed by EPR, our result
seems to lend support to the geometric nature of quantum entanglement. In
concluding remarks, we discuss very briefly a bold conjecture in physics
aiming to unify general relativity with quantum mechanics, namely, ER=EPR.
\end{abstract}

\section{Introduction}

Quantum entanglement has developed from a mere intellectual curiosity \cite%
{EPR} of the fundamental structure of quantum mechanics\footnote{%
Note that although violation of Bell's inequality theorem asserts the
nonlocality of quantum mechanics, the EPR inquiry is still not resolved,
i.e., what is still left unanswered is the mysterious 'link' between qubits
corresponding to our model \textit{via} a "see-saw" or mechanical inverter
representation of the Pauli-matrix inverter, $\sigma _{x}$.
\par
{}} to become an important and practical resource for quantum information
processing in the evolving theory of quantum information and ultra-fast
massively-parallel computing. Here, we give a new perspective of quantum
teleportation and superdense coding employing discrete phase-space physics,
superposition principle, and Hadamard transform, coupled with ICL
diagrammatic techniques \cite{entmeasure,entbasis}.

We first discuss the discrete phase-space formalism for entangled qubits. In
the discrete phase space formalism, entangled qubits are \textit{Bloch
function} states \cite{trhn, book}. From the general relation, we see that
we can construct prime-number dimensional spaces, besides the entangled
basis states, that are connected by Hadamard transform. In conventional
superdense coding quantum circuits, using controlled-not gates and Hadamard
transforms, these other \textit{Bloch function} states has been useful. In
our ICL diagrammatic techniques, we find no need to use these other \textit{%
Bloch function} states besides the entangled Bell basis states.

\section{Computational basis or \textit{Wannier} states}

\begin{figure}[hbt!]
\centering
\includegraphics[width=5.3039in]{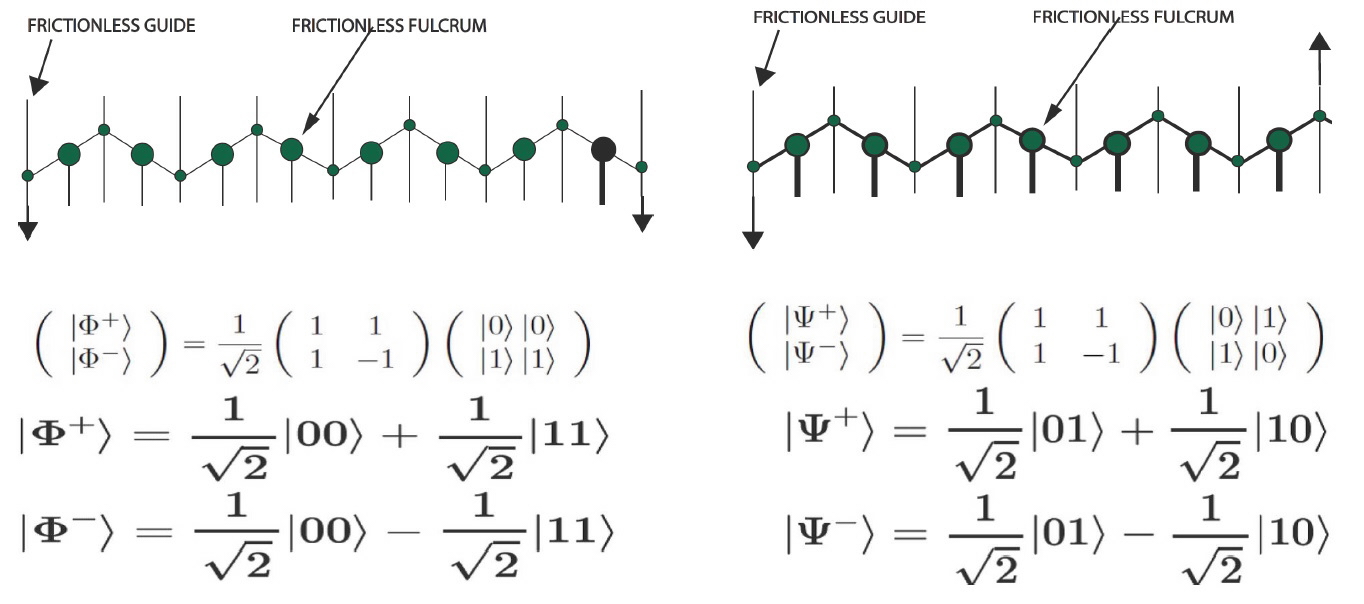}
\caption{Physical diagrammatic model of \textquotedblleft
triplet\textquotedblright\ (left) and singlet (right) entanglement. By
construction, each diagram is viewed as a two-state system, respectively.
The actual physical implementation of the chain of inverters may need
frictionles male/female sliding tube coupling for large-angle swing, but
this is beside the point. We assume a rigid coupling model for simultaneity
of events at both ends.}
\label{figA}
\end{figure}

Consider the four Wannier states,%
\begin{eqnarray}
\left\vert R_{0}\right\rangle &=&\left\vert 0\right\rangle _{1}\left\vert
0\right\rangle _{2}  \label{eq1} \\
\left\vert R_{1}\right\rangle &=&\left\vert 0\right\rangle _{1}\left\vert
1\right\rangle _{2}  \label{eq2} \\
\left\vert R_{2}\right\rangle &=&\left\vert 1\right\rangle _{1}\left\vert
0\right\rangle _{2}  \label{eq3} \\
\left\vert R_{3}\right\rangle &=&\left\vert 1\right\rangle _{1}\left\vert
1\right\rangle _{2}  \label{eq4} \\
\left\vert R_{4}\right\rangle &=&\left\vert R_{0}\right\rangle =\left\vert
0\right\rangle _{1}\left\vert 0\right\rangle _{2}  \label{eq5}
\end{eqnarray}%
Equation (\ref{eq5}) is the Born-von Karman boundary condition. Because the
number of "\textit{site"} states are $4$-even, mathematically this do not
represent a prime number which allows for finite field analyses. We seek
out, from the general discrete Fourier transform relation, reduced spaces
that are represented by prime number of sites. Obviously, the number $4$ is
a product of two prime numbers $2$. So we seek $2$-$D$ spaces connected by
Hadamard transform, from the general discrete Fourier transform relations.
We have the Fourier transformed states,%
\begin{equation}
\left\vert B_{k}\right\rangle =\frac{1}{\sqrt{N}}\sum_{R_{i},\
i=0...3}e^{ik\cdot R_{i}}\left\vert R_{i}\right\rangle  \label{eq6}
\end{equation}%
and the inverse transformation%
\begin{equation}
\left\vert R_{i}\right\rangle =\frac{1}{\sqrt{N}}\sum_{k_{j},\
i=0...3}e^{-ik_{j}\cdot R_{i}}\left\vert B_{k_{j}}\right\rangle  \label{eq7}
\end{equation}%
where%
\begin{eqnarray}
k_{n} &=&\frac{2\pi }{4}n\text{, \ }n=0,1,...3  \label{eq8} \\
k_{0} &=&0,\ k_{1}=\frac{\pi }{2},\ k_{2}=\pi ,\ k_{3}=\frac{3\pi }{2}
\label{eq9}
\end{eqnarray}%
\begin{equation}
e^{ik\cdot R}=\left( 
\begin{array}{cccc}
1 & 1 & 1 & 1 \\ 
1 & i & -1 & -i \\ 
1 & -1 & 1 & -1 \\ 
1 & -i & -1 & i%
\end{array}%
\right)  \label{eq10}
\end{equation}%
From Eq. (\ref{eq6}), we have%
\begin{equation}
\left\vert B_{k}\right\rangle =\frac{1}{\sqrt{N}}\left( 
\begin{array}{cccc}
1 & 1 & 1 & 1 \\ 
1 & i & -1 & -i \\ 
1 & -1 & 1 & -1 \\ 
1 & -i & -1 & i%
\end{array}%
\right) \left( 
\begin{array}{c}
\left\vert 0\right\rangle \left\vert 0\right\rangle \\ 
\left\vert 0\right\rangle \left\vert 1\right\rangle \\ 
\left\vert 1\right\rangle \left\vert 0\right\rangle \\ 
\left\vert 1\right\rangle \left\vert 1\right\rangle%
\end{array}%
\right)  \label{eq11}
\end{equation}%
For the inverse transformationfunction, we have,%
\begin{eqnarray}
e^{-ik\cdot R} &=&\left( 
\begin{array}{cccc}
e^{i0\cdot 0} & e^{i0\cdot 1} & e^{i0\cdot 2} & e^{i0\cdot 3} \\ 
e^{i0\cdot 1} & e^{-i\frac{\pi }{2}\cdot 1} & e^{-i\pi \cdot 1} & e^{-i\frac{%
3\pi }{2}\cdot 1} \\ 
e^{i0\cdot 2} & e^{-i\frac{\pi }{2}\cdot 2} & e^{-i\pi \cdot 2} & e^{-i\frac{%
3\pi }{2}\cdot 2} \\ 
e^{i0\cdot 3} & e^{-i\frac{\pi }{2}\cdot 3} & e^{-i\pi \cdot 3} & e^{-i\frac{%
3\pi }{2}\cdot 3}%
\end{array}%
\right)  \label{eq13} \\
&=&\left( 
\begin{array}{cccc}
1 & 1 & 1 & 1 \\ 
1 & -i & -1 & i \\ 
1 & -1 & 1 & -1 \\ 
1 & i & -1 & -i%
\end{array}%
\right)  \label{eq14}
\end{eqnarray}%
\begin{equation}
\left( e^{ik\cdot R}\right) \left( e^{-ik\cdot R}\right) =\frac{1}{4}\left( 
\begin{array}{cccc}
1 & 1 & 1 & 1 \\ 
1 & i & -1 & -i \\ 
1 & -1 & 1 & -1 \\ 
1 & -i & -1 & i%
\end{array}%
\right) \left( 
\begin{array}{cccc}
1 & 1 & 1 & 1 \\ 
1 & -i & -1 & i \\ 
1 & -1 & 1 & -1 \\ 
1 & i & -1 & -i%
\end{array}%
\right)  \label{eq15}
\end{equation}%
\begin{equation}
\left( e^{ik\cdot R}\right) \left( e^{-ik\cdot R}\right) =\left( 
\begin{array}{cccc}
1 & 0 & 0 & 0 \\ 
0 & 1 & 0 & 0 \\ 
0 & 0 & 1 & 0 \\ 
0 & 0 & 0 & 1%
\end{array}%
\right) =I_{4}  \label{eq17}
\end{equation}%
From Eq.(\ref{eq7}), we have,%
\begin{equation}
\left( 
\begin{array}{c}
\left\vert 0\right\rangle \left\vert 0\right\rangle \\ 
\left\vert 0\right\rangle \left\vert 1\right\rangle \\ 
\left\vert 1\right\rangle \left\vert 0\right\rangle \\ 
\left\vert 1\right\rangle \left\vert 1\right\rangle%
\end{array}%
\right) =\left( 
\begin{array}{cccc}
1 & 1 & 1 & 1 \\ 
1 & -i & -1 & i \\ 
1 & -1 & 1 & -1 \\ 
1 & i & -1 & -i%
\end{array}%
\right) \left( 
\begin{array}{c}
\left\vert B_{0}\right\rangle \\ 
\left\vert B_{\frac{\pi }{2}}\right\rangle \\ 
\left\vert B_{\pi }\right\rangle \\ 
\left\vert B_{\frac{3\pi }{2}}\right\rangle%
\end{array}%
\right)  \label{eq19}
\end{equation}%
where, from Eq.(\ref{eq6}),%
\begin{equation}
\left( 
\begin{array}{c}
\left\vert B_{0}\right\rangle \\ 
\left\vert B_{\frac{\pi }{2}}\right\rangle \\ 
\left\vert B_{\pi }\right\rangle \\ 
\left\vert B_{\frac{3\pi }{2}}\right\rangle%
\end{array}%
\right) =\frac{1}{2}\left( 
\begin{array}{cccc}
1 & 1 & 1 & 1 \\ 
1 & i & -1 & -i \\ 
1 & -1 & 1 & -1 \\ 
1 & -i & -1 & i%
\end{array}%
\right) \left( 
\begin{array}{c}
\left\vert 0\right\rangle _{1}\left\vert 0\right\rangle _{2} \\ 
\left\vert 0\right\rangle _{1}\left\vert 1\right\rangle _{2} \\ 
\left\vert 1\right\rangle _{1}\left\vert 0\right\rangle _{2} \\ 
\left\vert 1\right\rangle _{1}\left\vert 1\right\rangle _{2}%
\end{array}%
\right)  \label{eq20}
\end{equation}
However, as depicted in our ICL model, entanglement can only occurs for two
states, namely, either%
\begin{equation}
\left( 
\begin{array}{c}
\left\vert 0\right\rangle _{1}\left\vert 0\right\rangle _{2} \\ 
0 \\ 
0 \\ 
\left\vert 1\right\rangle _{1}\left\vert 1\right\rangle _{2}%
\end{array}%
\right) \text{ or }\left( 
\begin{array}{c}
0 \\ 
\left\vert 0\right\rangle _{1}\left\vert 1\right\rangle _{2} \\ 
\left\vert 1\right\rangle _{1}\left\vert 0\right\rangle _{2} \\ 
0%
\end{array}%
\right) \text{.}  \label{eq23}
\end{equation}%
These are faithfully represented by our ICL model, Fig. \ref{figA}.

So, we are only interested in the Bloch functions, coming from $\left\vert
B_{0}\right\rangle $ and $\left\vert B_{2}\right\rangle $ having Hadamard
Fourier transformation from two site states. We have the remaining matrix
entries,%
\begin{equation}
\left( 
\begin{array}{c}
\left\vert B_{0}\right\rangle =\Phi ^{+} \\ 
0 \\ 
\left\vert B_{2}\right\rangle =\Phi ^{-} \\ 
0%
\end{array}%
\right) =\frac{1}{\sqrt{2}}\left( 
\begin{array}{cccc}
1 & 0 & 0 & 1 \\ 
0 & 0 & 0 & 0 \\ 
1 & 0 & 0 & -1 \\ 
0 & 0 & 0 & 0%
\end{array}%
\right) \left( 
\begin{array}{c}
\left\vert 0\right\rangle _{1}\left\vert 0\right\rangle _{2} \\ 
\left\vert 0\right\rangle _{1}\left\vert 1\right\rangle _{2} \\ 
\left\vert 1\right\rangle _{1}\left\vert 0\right\rangle _{2} \\ 
\left\vert 1\right\rangle _{1}\left\vert 1\right\rangle _{2}%
\end{array}%
\right)   \label{eq24}
\end{equation}%
where the $2$-$D$ transformation matrix is derived from the general
relation, 
\begin{equation}
\left( 
\begin{array}{c}
\left\vert B_{0}\right\rangle  \\ 
\left\vert B_{1}\right\rangle  \\ 
\left\vert B_{2}\right\rangle  \\ 
\left\vert B_{3}\right\rangle 
\end{array}%
\right) =\frac{1}{2}\left( 
\begin{array}{cccc}
\delta  & 1 & 1 & \delta  \\ 
1 & i & -1 & -i \\ 
\delta  & -1 & 1 & -\delta  \\ 
1 & -i & -1 & i%
\end{array}%
\right) \left( 
\begin{array}{c}
\left\vert 0\right\rangle _{1}\left\vert 0\right\rangle _{2} \\ 
\left\vert 0\right\rangle _{1}\left\vert 1\right\rangle _{2} \\ 
\left\vert 1\right\rangle _{1}\left\vert 0\right\rangle _{2} \\ 
\left\vert 1\right\rangle _{1}\left\vert 1\right\rangle _{2}%
\end{array}%
\right)   \label{eq25}
\end{equation}%
where all matrix entries are cross out except the $\delta $ entries in Eq. (%
\ref{eq25}). Therefore by contracting to prime number space, we have, using
only the $\delta $ entries of the transformation matrix,%
\begin{equation}
\left( 
\begin{array}{c}
\left\vert \Phi ^{+}\right\rangle  \\ 
\left\vert \Phi ^{-}\right\rangle 
\end{array}%
\right) =\left( 
\begin{array}{c}
\left\vert B_{0}\right\rangle  \\ 
\left\vert B_{2}\right\rangle 
\end{array}%
\right) =\frac{1}{\sqrt{2}}\left( 
\begin{array}{cc}
1 & 1 \\ 
1 & -1%
\end{array}%
\right) \left( 
\begin{array}{c}
\left\vert 0\right\rangle _{1}\left\vert 0\right\rangle _{2} \\ 
\left\vert 1\right\rangle _{1}\left\vert 1\right\rangle _{2}%
\end{array}%
\right)   \label{eq26}
\end{equation}%
or with other contraction of $\left\vert B_{0}\right\rangle $ and $%
\left\vert B_{2}\right\rangle $ retaining the transformation, 
\begin{equation}
\left( 
\begin{array}{c}
\left\vert B_{0}\right\rangle =\Psi ^{+} \\ 
0 \\ 
\left\vert B_{2}\right\rangle =-\Psi ^{-} \\ 
0%
\end{array}%
\right) =\frac{1}{\sqrt{2}}\left( 
\begin{array}{cccc}
0 & 1 & 1 & 0 \\ 
0 & 0 & 0 & 0 \\ 
0 & -1 & 1 & 0 \\ 
0 & 0 & 0 & 0%
\end{array}%
\right) \left( 
\begin{array}{c}
\left\vert 0\right\rangle _{1}\left\vert 0\right\rangle _{2} \\ 
\left\vert 0\right\rangle _{1}\left\vert 1\right\rangle _{2} \\ 
\left\vert 1\right\rangle _{1}\left\vert 0\right\rangle _{2} \\ 
\left\vert 1\right\rangle _{1}\left\vert 1\right\rangle _{2}%
\end{array}%
\right)   \label{eq27}
\end{equation}%
yielding $2$-$D$ transformation matrix%
\begin{equation}
\left( 
\begin{array}{c}
\Psi ^{+} \\ 
-\Psi ^{-}%
\end{array}%
\right) =\left( 
\begin{array}{c}
\left\vert B_{0}\right\rangle  \\ 
\left\vert B_{2}\right\rangle 
\end{array}%
\right) =\frac{1}{\sqrt{2}}\left( 
\begin{array}{cc}
1 & 1 \\ 
-1 & 1%
\end{array}%
\right) \left( 
\begin{array}{c}
\left\vert 0\right\rangle _{1}\left\vert 1\right\rangle _{2} \\ 
\left\vert 1\right\rangle _{1}\left\vert 0\right\rangle _{2}%
\end{array}%
\right)   \label{eq28}
\end{equation}%
\begin{equation}
\left( 
\begin{array}{c}
\Psi ^{+} \\ 
-\Psi ^{-}%
\end{array}%
\right) \Longrightarrow \left( 
\begin{array}{c}
\Psi ^{+} \\ 
\Psi ^{-}%
\end{array}%
\right) =\frac{1}{\sqrt{2}}\left( 
\begin{array}{cc}
1 & 1 \\ 
1 & -1%
\end{array}%
\right) \left( 
\begin{array}{c}
\left\vert 0\right\rangle _{1}\left\vert 1\right\rangle _{2} \\ 
\left\vert 1\right\rangle _{1}\left\vert 0\right\rangle _{2}%
\end{array}%
\right)   \label{eq29}
\end{equation}%
where $\frac{1}{\sqrt{2}}$ becomes the proper normalization factor for the
reduced $2-D$ space.

Of course, besides the above entangled Bell basis states, there are other
interesting Bloch functions with Hadamard transformations, by selectively
crossing out of the matrix entries of Eq. (\ref{eq25}), such as, 
\begin{eqnarray}
&&\left( 
\begin{array}{c}
\left\vert H_{0}\right\rangle \\ 
\left\vert H_{1}\right\rangle%
\end{array}%
\right) \text{{\tiny derived from}}\left( 
\begin{array}{c}
\left\vert B_{0}\right\rangle \\ 
\left\vert B_{1}\right\rangle%
\end{array}%
\right)  \nonumber \\
&=&\frac{1}{\sqrt{2}}\left( 
\begin{array}{cc}
1 & 1 \\ 
1 & -1%
\end{array}%
\right) \left( 
\begin{array}{c}
\left\vert 0\right\rangle _{1}\left\vert 0\right\rangle _{2} \\ 
\left\vert 1\right\rangle _{1}\left\vert 0\right\rangle _{2}%
\end{array}%
\right)  \label{eq30}
\end{eqnarray}%
\begin{eqnarray}
&&\left( 
\begin{array}{c}
\left\vert H_{2}\right\rangle \\ 
\left\vert H_{3}\right\rangle%
\end{array}%
\right) \text{{\tiny derived from}}\left( 
\begin{array}{c}
\left\vert B_{0}\right\rangle \\ 
\left\vert B_{2}\right\rangle%
\end{array}%
\right)  \nonumber \\
&=&\frac{1}{\sqrt{2}}\left( 
\begin{array}{cc}
1 & 1 \\ 
1 & -1%
\end{array}%
\right) \left( 
\begin{array}{c}
\left\vert 0\right\rangle _{1}\left\vert 0\right\rangle _{2} \\ 
\left\vert 0\right\rangle _{1}\left\vert 1\right\rangle _{2}%
\end{array}%
\right)  \label{eq31}
\end{eqnarray}%
\begin{eqnarray}
&&\left( 
\begin{array}{c}
\left\vert H_{0}\right\rangle \\ 
\left\vert H_{1}\right\rangle%
\end{array}%
\right) \text{{\tiny derived from}}\left( 
\begin{array}{c}
\left\vert B_{0}\right\rangle \\ 
\left\vert B_{3}\right\rangle%
\end{array}%
\right)  \nonumber \\
&=&\frac{1}{\sqrt{2}}\left( 
\begin{array}{cc}
1 & 1 \\ 
1 & -1%
\end{array}%
\right) \left( 
\begin{array}{c}
\left\vert 0\right\rangle _{1}\left\vert 0\right\rangle _{2} \\ 
\left\vert 1\right\rangle _{1}\left\vert 0\right\rangle _{2}%
\end{array}%
\right)  \label{eq32}
\end{eqnarray}%
\begin{eqnarray}
&&\left( 
\begin{array}{c}
\left\vert H_{0}\right\rangle \\ 
\left\vert H_{1}\right\rangle%
\end{array}%
\right) \text{{\tiny derived from} }\left( 
\begin{array}{c}
\left\vert B_{2}\right\rangle \\ 
\left\vert B_{3}\right\rangle%
\end{array}%
\right)  \nonumber \\
&=&\frac{1}{\sqrt{2}}\left( 
\begin{array}{cc}
1 & 1 \\ 
1 & -1%
\end{array}%
\right) \left( 
\begin{array}{c}
\left\vert 0\right\rangle _{1}\left\vert 0\right\rangle _{2} \\ 
\left\vert 1\right\rangle _{1}\left\vert 0\right\rangle _{2}%
\end{array}%
\right)  \label{eq33}
\end{eqnarray}%
\begin{eqnarray}
&&\left( 
\begin{array}{c}
\left\vert H_{4}\right\rangle \\ 
\left\vert H_{5}\right\rangle%
\end{array}%
\right) \text{{\tiny derived from}}\left( 
\begin{array}{c}
\left\vert B_{0}\right\rangle \\ 
\left\vert B_{3}\right\rangle%
\end{array}%
\right)  \nonumber \\
&=&\left( 
\begin{array}{cc}
1 & 1 \\ 
1 & -1%
\end{array}%
\right) \left( 
\begin{array}{c}
\left\vert 1\right\rangle _{1}\left\vert 0\right\rangle _{2} \\ 
\left\vert 1\right\rangle _{1}\left\vert 1\right\rangle _{2}%
\end{array}%
\right)  \label{eq34}
\end{eqnarray}%
Equations (\ref{eq30})-(\ref{eq34}) cannot be represented in terms of our
ICL model, and therefore do not represent entangled states. Thus, \textit{%
only entangled states} are faithfully representable in terms of our ICL
model. Clearly, the entanglement of two bare qubits is divided into two
orthogonal spaces of triplet\footnote{%
The use of the term \textquotedblleft triplet\textquotedblright\ is actually
a misnomer here since the entangled system is not free to assume a singlet
or zero spin state. Thus, this term is used here only as a label} and
singlet entanglement states in Fig.\ref{figA}. These two spaces are
transformed by the transition-matrix, $\sigma _{x}$.

Mathematically, the inverter-chain link model of entanglement may be
formulated as a series of $\sigma _{x}$ operations, represented by physical
inverters or \textit{see-saw}'s. Assume at first that there are two
locally-entangled qubits $A$ and $B$ in either singlet or triplet state,
with singlet joined with one \textit{see-saw} ($\sigma _{x}$) or triplet
joined by two \textit{see-saw}'s ($\sigma _{x}\otimes \sigma _{x}$). Then
using the unitary single-qubit operation, $\sigma _{x}$, on one of the two
qubits will result in an additional extension of a \textit{i}%
nverter-chain-linked entangled two qubits, either $\Phi ^{+}$ or $\Psi ^{+}$%
, depending on the initial singlet or triplet state. Using a series of $%
\sigma _{x}$ operations will then yield a \textit{physically longer}
inverter-chain link between the two entangled qubits. A series of odd number
of $\sigma _{x}$ operations will result in eventual inversion of one of the
qubit, whereas an even number of $\sigma _{x}$ operations is equivalent to
an identify operation of one of the qubit, although the inverter-chain link
is always extended by one \textit{see-saw} with each $\sigma _{x}$ operation.

This arbitrary extension of the $\sigma _{x}$ inverter-chain is the reason
why $A$\textit{\ and }$B$\textit{\ can be separated at arbitrarily large
distances }while still being entangled, providing the \textit{mysterious}
link raised by the EPR \cite{EPR}. Thus, the inverter-chain link arbitrary
extension defines local and nonlocal entanglements. It should be emphasized
that our inverter-chain link model is a physical \textit{representation} of $%
\sigma _{x}$ operations. Indeed, a more fundamental geometric representation
of the \textit{mysterious} link raised by the EPR \cite{EPR} will unify
general relativity and quantum mechanics and has been challenging
theoretical physicists for decades.

Clearly, a \textit{single qubit }$\sigma _{x}$\textit{\ operation is
equivalent to extending the inverter- link chain link by one physical
inverter,} which basically serves as a single-qubit unitary transformation
from $\Phi ^{+}$ to $\Psi ^{+}$ and \textit{vice vers}a. Note in all these
entangled cases, the following unitary transformations, namely, $\sigma _{x}$
and $\sigma _{z}$ on a qubit partner generates the other entangled states
and its phase $\sigma _{z}$-transformed entangled partner, namely,%
\begin{equation}
\sigma _{x}\left\vert \Phi ^{+}\right\rangle =\left\vert \Psi
^{+}\right\rangle  \label{eq37}
\end{equation}%
\begin{eqnarray}
\sigma _{z}\left\vert \Phi ^{+}\right\rangle &=&\left\vert \Phi
^{-}\right\rangle  \label{eq35} \\
\sigma _{z}\left\vert \Psi ^{+}\right\rangle &=&\left\vert \Psi
^{-}\right\rangle  \label{eq36}
\end{eqnarray}%
where in Eqs. (\ref{eq37}) - (\ref{eq36}), the Pauli operators operates only
on single qubit.The inverse Hadamard transformation yields the \textit{%
Wannier states, i.e., } 
\begin{eqnarray}
\frac{1}{\sqrt{2}}\left( \left\vert \Phi ^{+}\right\rangle +\left\vert \Phi
^{-}\right\rangle \right) &=&\left\vert 0\right\rangle _{1}\left\vert
0\right\rangle _{2}  \label{eq38} \\
\frac{1}{\sqrt{2}}\left( \left\vert \Phi ^{+}\right\rangle -\left\vert \Phi
^{-}\right\rangle \right) &=&\left\vert 1\right\rangle _{1}\left\vert
1\right\rangle _{2}  \label{eq39}
\end{eqnarray}%
\begin{eqnarray}
\frac{1}{\sqrt{2}}\left( \left\vert \Psi ^{+}\right\rangle +\left\vert \Psi
^{-}\right\rangle \right) &=&\left\vert 0\right\rangle _{1}\left\vert
1\right\rangle _{2}  \label{eq40} \\
\frac{1}{\sqrt{2}}\left( \left\vert \Psi ^{+}\right\rangle -\left\vert \Psi
^{-}\right\rangle \right) &=&\left\vert 1\right\rangle _{1}\left\vert
0\right\rangle _{2}  \label{eq41}
\end{eqnarray}%
Equations (\ref{eq38}) - (\ref{eq41}) are crucial in our ICL diagrammatic
analysis of superdense coding.

For the unentangled superpositions or other \textit{Bloch function states},
we also have%
\begin{eqnarray}
\frac{1}{\sqrt{2}}\left( \left\vert H_{0}\right\rangle +\left\vert
H_{1}\right\rangle \right)  &=&\left\vert 0\right\rangle _{1}\left\vert
0\right\rangle _{2}  \label{eq42} \\
\frac{1}{\sqrt{2}}\left( \left\vert H_{0}\right\rangle -\left\vert
H_{1}\right\rangle \right)  &=&\left\vert 1\right\rangle _{1}\left\vert
0\right\rangle _{2}  \label{eq43}
\end{eqnarray}%
\begin{eqnarray}
\frac{1}{\sqrt{2}}\left( \left\vert H_{2}\right\rangle +\left\vert
H_{3}\right\rangle \right)  &=&\left\vert 0\right\rangle _{1}\left\vert
0\right\rangle _{2}  \label{eq44} \\
\frac{1}{\sqrt{2}}\left( \left\vert H_{2}\right\rangle -\left\vert
H_{3}\right\rangle \right)  &=&\left\vert 0\right\rangle _{1}\left\vert
1\right\rangle _{2}  \label{q45}
\end{eqnarray}%
\begin{eqnarray}
\frac{1}{\sqrt{2}}\left( \left\vert H_{4}\right\rangle +\left\vert
H_{5}\right\rangle \right)  &=&\left\vert 1\right\rangle _{1}\left\vert
0\right\rangle _{2}  \label{eq46} \\
\frac{1}{\sqrt{2}}\left( \left\vert H_{4}\right\rangle -\left\vert
H_{5}\right\rangle \right)  &=&\left\vert 1\right\rangle _{1}\left\vert
1\right\rangle _{2}  \label{eq47}
\end{eqnarray}%
Equations (\ref{eq42}) - (\ref{eq47}) has been useful in the conventional
quantum circuit diagrams, using controlled gates and Hadamard transforms,
for superdense coding. Again, in our ICL diagrams we find no need for these
other unentangled \textit{Bloch function} states to elucidate the
fundamental concept.

\subsection{Unitary transformation of qubits}

In matrix representation, a single qubit is represented by a matrix,%
\begin{eqnarray}
\left\vert 0\right\rangle &\equiv &\left( 
\begin{array}{c}
\left\vert 0\right\rangle \\ 
0%
\end{array}%
\right) \Longrightarrow \left( 
\begin{array}{c}
1 \\ 
0%
\end{array}%
\right)  \label{eq48} \\
\left\vert 1\right\rangle &\equiv &\left( 
\begin{array}{c}
0 \\ 
\left\vert 1\right\rangle%
\end{array}%
\right) \Longrightarrow \left( 
\begin{array}{c}
0 \\ 
1%
\end{array}%
\right)  \label{eq49}
\end{eqnarray}%
The unitary transformations of a qubit are the $\sigma _{x},$\ $\sigma _{z},$%
\ and their combinations, where the $\sigma _{x},$\ and $\sigma _{z}$ are
the Pauli matrices. $\sigma _{x}$ acts as an inverter, whereas $\sigma _{z}$
acts as a phase operator, when these operators are acting on the \textit{%
Wannier} states. Thus, we 
\begin{eqnarray}
\sigma _{x}\left\vert 0\right\rangle &=&\left( 
\begin{array}{c}
0 \\ 
1%
\end{array}%
\right) =\left\vert 1\right\rangle  \label{eq50} \\
\sigma _{x}\left\vert 1\right\rangle &=&\left( 
\begin{array}{c}
1 \\ 
0%
\end{array}%
\right) =\left\vert 0\right\rangle  \label{q51} \\
\sigma _{z}\left\vert 0\right\rangle &=&\left( 
\begin{array}{c}
1 \\ 
0%
\end{array}%
\right) =\left\vert 0\right\rangle  \label{eq52} \\
\sigma _{z}\left\vert 1\right\rangle &=&\left( 
\begin{array}{c}
0 \\ 
-1%
\end{array}%
\right) =-\left\vert 1\right\rangle  \label{eq53}
\end{eqnarray}

\section{ICL implementation of quantum teleportation}

The diagrammatic perspective relies on a couple of principles, namely, (1)
that unitary transformations of a partner qubit generate all other possible
entanglements from a given entangled qubits, and (2) that quantum
information or quantum configurations (entropy) is conserved in quantum
teleportation. The first principle allows us to examine all possible
scenarios of quantum teleportation. Coupled with classical communication
(CC) protocol, the second principle conserved the configurations of the
system.

\begin{figure}[tbh]
\centering
\includegraphics[width=5.8747in]{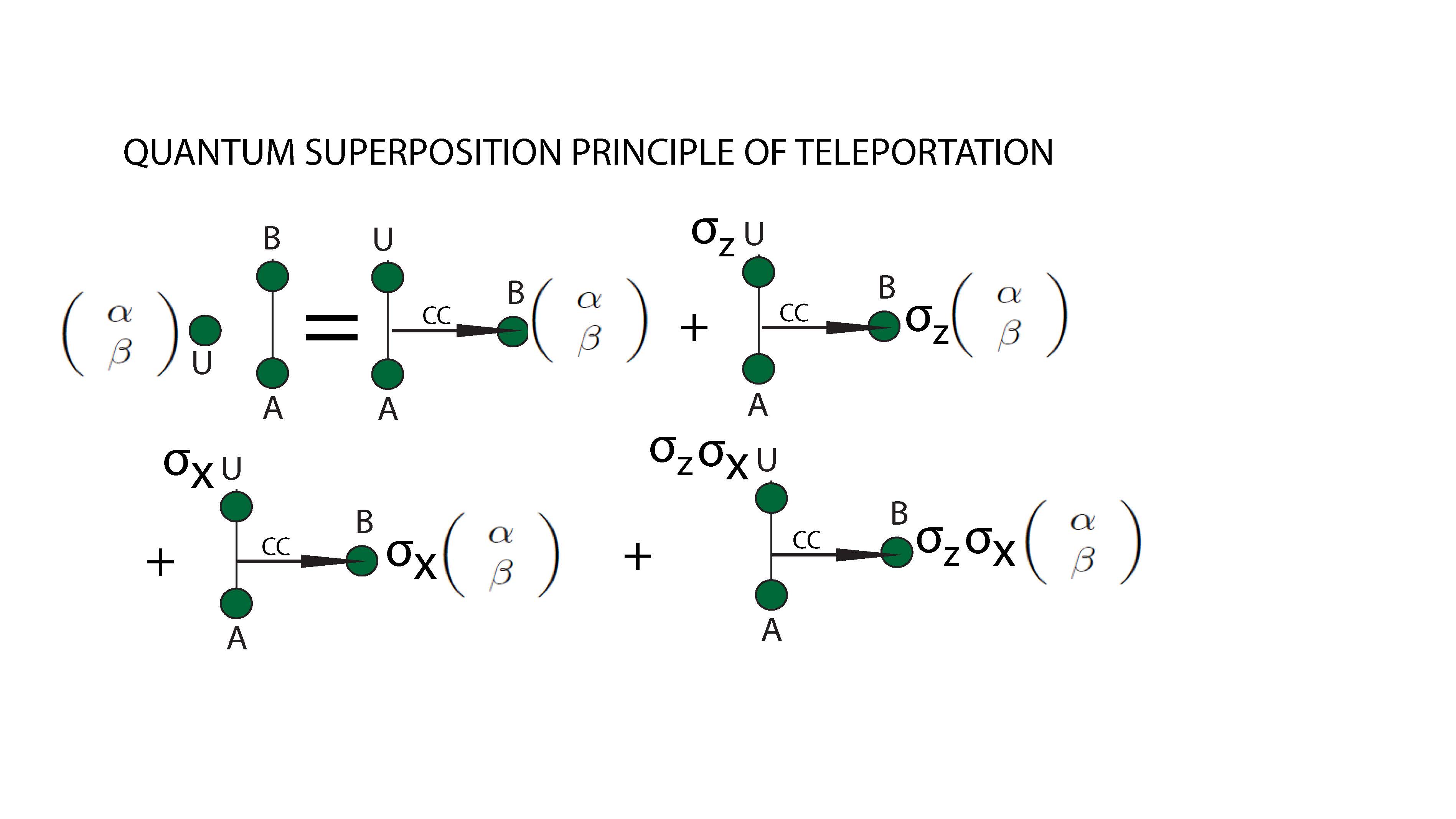}
\caption{Explicit superposition principle in quantum teleportation, depicted
in terms of our ICL model of quantum entanglement. The term 'cc' means
classical communication.}
\label{fig1}
\end{figure}

The virtue of the diagrammatic perspective is that it simplifies the process
and gives a very intuitive understanding of quantum teleportation. Let Alice
and Bob share a maximally entangled state initially,%
\begin{equation}
\left\vert \Phi ^{+}\right\rangle _{AB}=\frac{1}{\sqrt{2}}\left( \left\vert
0\right\rangle _{A}\left\vert 0\right\rangle _{B}+\left\vert 1\right\rangle
_{A}\left\vert 1\right\rangle _{B}\right)  \label{eq54}
\end{equation}%
So, Alice has two particles ($U$, the one she wants to teleport, and $A$,
one of the entangled pair), and Bob has the other entangled-pair particle, $%
B $. Mathematically, we have,%
\begin{eqnarray}
&&\left\vert \psi \right\rangle _{U}\otimes \left\vert \Phi
^{+}\right\rangle _{AB}=C_{B}^{\left( +\right) }\left\vert \Phi
^{+}\right\rangle _{UA}+C_{B}^{\left( -\right) }\left\vert \Phi
^{-}\right\rangle _{UA}+D_{B}^{\left( +\right) }\left\vert \Psi
^{+}\right\rangle _{UA}...  \nonumber \\
&=&\frac{1}{2}\left\vert \Phi ^{+}\right\rangle _{UA}\ \otimes \psi
_{_{B}}^{\left( 0\right) }+\frac{1}{2}\left\vert \Phi ^{-}\right\rangle
_{UA}\ \otimes \psi _{_{B}}^{\left( 1\right) }  \nonumber \\
&&+\frac{1}{2}\left\vert \Psi ^{+}\right\rangle _{UA}\ \otimes \psi
_{_{B}}^{\left( 2\right) }+\frac{1}{2}\left\vert \Psi ^{-}\right\rangle
_{UA}\otimes \psi _{_{B}}^{\left( 3\right) }  \label{eq57}
\end{eqnarray}%
where, $\psi _{_{B}}^{\left( i\right) }$, \ $\left( \ i=1,2,3\right) $ are
the coefficients of the expansion in terms of complete \textit{entangled
basis states} of $A$ with $U$, namely, $\left\vert \Phi ^{+}\right\rangle
_{UA}$, $\left\vert \Phi ^{-}\right\rangle _{UA}$, $\left\vert \Psi
^{+}\right\rangle _{UA}$, $\left\vert \Psi ^{-}\right\rangle _{UA}$. More
explicitly, we have,%
\begin{eqnarray}
\left( 
\begin{array}{c}
\alpha \\ 
\beta%
\end{array}%
\right) _{U}\otimes \left\vert \Phi ^{+}\right\rangle _{AB} &=&\frac{1}{2}%
\left\vert \Phi ^{+}\right\rangle _{UA}\left( 
\begin{array}{c}
\alpha \\ 
\beta%
\end{array}%
\right)  \nonumber \\
&&+\frac{1}{2}\left\vert \Phi ^{-}\right\rangle _{UA}\left( 
\begin{array}{cc}
1 & 0 \\ 
0 & -1%
\end{array}%
\right) \left( 
\begin{array}{c}
\alpha \\ 
\beta%
\end{array}%
\right)  \nonumber \\
&&+\frac{1}{2}\left\vert \Psi ^{+}\right\rangle _{UA}\left( 
\begin{array}{cc}
0 & 1 \\ 
1 & 0%
\end{array}%
\right) \left( 
\begin{array}{c}
\alpha \\ 
\beta%
\end{array}%
\right)  \nonumber \\
&&+\frac{1}{2}\left\vert \Psi ^{-}\right\rangle _{UA}\left( 
\begin{array}{cc}
0 & -1 \\ 
1 & 0%
\end{array}%
\right) \left( 
\begin{array}{c}
\alpha \\ 
\beta%
\end{array}%
\right) .  \label{eq58}
\end{eqnarray}%
The last line of Eq. (\ref{eq58}) differs only by a global sign factor with
the term, 
\[
+\frac{1}{2}\left\vert \Psi ^{-}\right\rangle _{UA}\left( 
\begin{array}{cc}
0 & 1 \\ 
-1 & 0%
\end{array}%
\right) \left( 
\begin{array}{c}
\alpha \\ 
\beta%
\end{array}%
\right) =+\frac{1}{2}\left\vert \Psi ^{-}\right\rangle _{UA}\ \sigma
_{z}\sigma _{x}\left( 
\begin{array}{c}
\alpha \\ 
\beta%
\end{array}%
\right) 
\]%
Thus, the situation looks exactly like the one shown in Fig.\ref{fig1}.

\section{ICL perspective of superdense coding}

The use of entanglement also plays an important resource in superdense
coding With Alice and Bob sharing an entangled state, two classical bits per
qubit can be transmitted. Superdense refers to this information capacity in
communication channels. Our diagrammatic approach greatly simplifies the
concept than that of the quantum circuit, controlled gate and Hadamard
tranformation, approach.

\begin{figure}[hbt!]
\centering
\includegraphics[width=5.6377in]{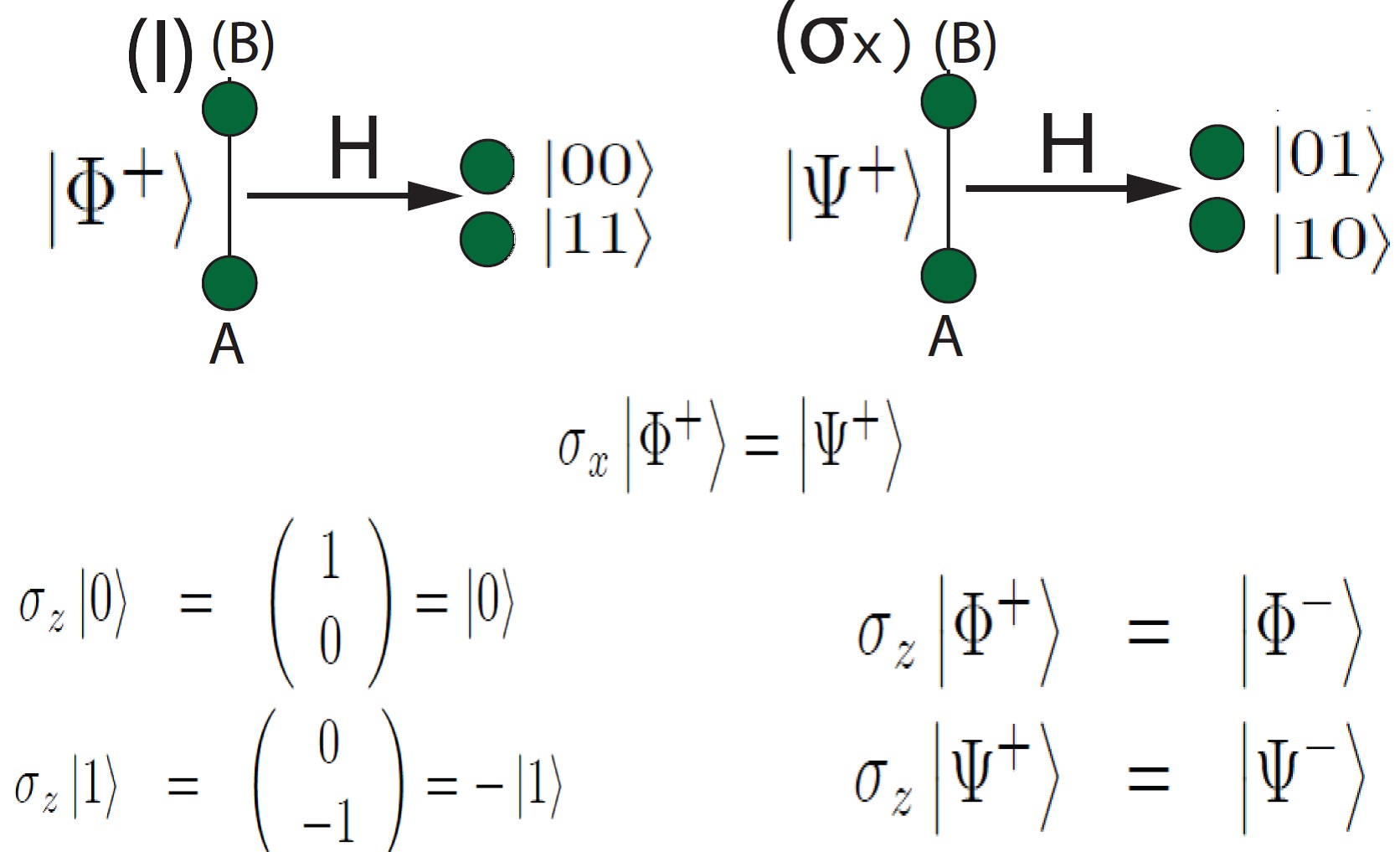}
\caption{ICL implementation of superdense coding. \textbf{H} is the Hadamard
transformation of a Bell basis and its phase-transformed partner.}
\label{fig2}
\end{figure}

Superdense coding is based on the observation that given some entangled Bell
basis state shared by Alice and Bob, there are local unitaries belonging to
either Alice or Bob which will map their shared Bell basis onto any of the
other Bell basis states apart from overall phase. Hadamard transformation, $%
H $, with its phase-transformed ($\sigma _{z}$) partner \textit{Bloch}
states then decode the resulting pair of \textit{Bloch} functions to their
computational basis or \textit{Wannier} function states. This is
schemtically illustrated in Fig.\ref{fig2}

\section{Concluding Remarks}

We have demonstrated that the discrete phase-space physics coupled with the
ICL implementation of entanglements have a much wider implications in
elucidating entanglement physics. For one thing, ICL has clearly shed more
light on the fundamental concept of concurrence, entropy of entanglement
formation, and entropic-distance measure as the fundamental characterization
of entanglement \cite{entmeasure}; thus, as a natural measures of
entanglement. The derivation of entangled basis states through the ICL
diagrams has also given a straightforward generalization of Bell basis
states for any number multi-partite entanglements \cite{entbasis}.

On the more fundamental question on the completeness of quantum physics, ICL
description of entanglement lends strong support on the significance of EPR
assertion that entanglements need a mysterious link, perhaps geometric,
which may call for extra dimensions. The apparently perfect scheme of the
ICL model in describing quantum entanglement seems to advocate the existence
of entangled spacetime-medium to take the place of the ICL in some unknown
forms. An interesting analogy is the bound/entangled pair of vortices and
antivortex on the surface of fluids which are really the ends of a vortex
chain (tube) forming a U-shape underneath the surface. For example, in $3$-$%
D $ the superfluid quantized vortices form a metastable closed ring or open
chain ending at the surface. A vortex chain with both ends ending at the
same surface appears as a bound vortex pair at the surface.

Thus, it seems extra dimensions are needed in spacetime to have a
fundamental theory of entanglement. The work of Ooguri \cite{vix,tokyo} and
collaborators shows that this quantum entanglement generates the extra
dimensions of the gravitational theory. "It appears that it seems possible
to generate a \textit{geometric} connection between entangled qubits, even
though there is no direct interaction between the two systems \cite%
{malcedona}. Furthermore, the structure of spacetime is proposed to be due
to the ghostly features of entanglement. Could it be that besides the
geometrical spacetime aspects of gravity, there is a purely quantum
mechanical aspect of spacetime geometry with extra dimensions that give rise
to entanglement?

A hint along this idea is also given by Malcedona \cite{malcedona} when he
stated that "One can consider, therefore, a pair of black holes where all
the microstates are \textquotedblleft entangled.\textquotedblright\ Namely,
if we observe one of the black holes in one particular microstate, then the
other has to be in exactly the same microstate. A pair of black holes in
this particular EPR entangled state would develop a wormhole, or
Einstein-Rosen bridge, connecting them through the \textit{inside}. The
geometry of this wormhole is given by the fully extended Schwarzschild
geometry. It is interesting that both wormholes and entanglement naively
appear to lead to a propagation of signals faster than light.

It was known that quantum entanglement is related to deep issues in the
unification of general relativity and quantum mechanics, such as the black
hole information paradox and the firewall paradox," says Hirosi Ooguri" \cite%
{tokyo}. Remarkably, all this can very simply be viewed as bound/entangled
pair of vortices and antivortex on the surface of fluids which are really
the ends of a vortex chain (vortex tube or "Einstein-Rosen bridge") forming
a U-shape chain underneath the surface or as a meeting two vertical funnels
if surface is theoretically bowed into U-shaped itself. This has led to the
proposition, ER=EPR conjecture in physics, stating that the EPR paradox
should lead to the unification of general relativity and quantum field
theory. Indeed, the ER = EPR conjecture \cite{suskind, verlinde} is the bold
statement that a large amount of entanglement between two localized regions
of space-time implies the existence of a \textit{geometric} connection
between two regions of space-time.


\begin{thebibliography}{99}
\bibitem{EPR} A. Einstein and B. Podolsky and N. Rosen, \textit{Can
Quantum-Mechanical Description of Physical Reality be Considered Complete?},
Phys. Rev. \textbf{47}, 777-780 (1937).

\bibitem{entmeasure} F.A. Buot, \textit{Perspective Chapter: On Entanglement
Measures -- Discrete Phase Space and Inverter-Chain Link Viewpoint},
https://www.intechopen.com/online-first/1162369

\bibitem{entbasis} F.A. Buot, A.R. Elnar, G. Maglasang, and C.M. Galon, 
\textit{A Mechanical Implementation and Diagrammatic Calculation of
Entangled Basis States}, https://arxiv.org/abs/2112.10291

\bibitem{trhn} F.A. Buot, \textit{Method for Calculating }$TrH^{n}$\textit{\
in Solid State Theory}, Phys. Rev., \textbf{B10}, 3700-3705(1974).

\bibitem{book} F.A. Buot. Nonequilibrium Quantum Transport Physics in
Nanosystem (World Scientific NJ, USA, 2009), and references therein

\bibitem{vix} G. Rajna, \textit{Spacetime is built by Quantum Entanglement},
https://vixra.org/pdf/1505.0206v1.pdf

\bibitem{tokyo} The University of Tokyo, \textit{How spacetime is built by
quantum entanglement},
https://phys.org/news/2015-05-spacetime-built-quantum-entanglement.html

\bibitem{malcedona} J. Maldacena, \textit{Entanglement and the Geometry of
Spacetime}, https://www.ias.edu/ideas/2013/maldacena-entanglement

\bibitem{suskind} J. Maldacena and L. Susskind, \textit{Cool horizons for
entangled black holes}, Fortsch. Phys. \textbf{61}, 781 (2013).

\bibitem{verlinde} E. Verlinde and H. Verlinde, \textit{A Conversation on ER
= EPR}, https://arxiv.org/pdf/2212.09389v1.pdf
\end{thebibliography}
\end{document}